\documentclass[preprint,showpacs]{revtex4}
\usepackage{mathrsfs}
\usepackage{amsfonts}

\usepackage{graphicx}
\usepackage{}% Include figure files
\usepackage{amsmath}% Include figure files
\usepackage{dcolumn}% Align table columns on decimal point
\usepackage{color}

\begin{document}
\title{Mechanically-mediated optical response in hybrid opto-electromechanical systems}

\author{Cheng Jiang$^{1}$}
\email{chengjiang_8402@163.com}
\author{Yuanshun Cui$^{1}$}
\author{Hongxiang Liu$^{1,2}$}
\author{Guibin Chen$^{1}$}
 \affiliation{
 $^1$ School of Physics and Electronic Electrical Engineering, Huaiyin Normal University, 111 West Chang Jiang Road, Huai'an 223001, China\\
 $^2$ School of Physics, Northeast Normal University, Changchun, 130024, China}
\date{\today}
\begin{abstract}
We theoretically investigate the analog of electromagnetically induced transparency, absorption and parametric amplification in a hybrid opto-electromechanical system consisting of an optical cavity and a microwave cavity coupled to a common nanomechanical resonator. When the two cavity modes are driven on their respective red sidebands by two pump beams and a probe beam is applied to the optical cavity to monitor the optical response of the hybrid system, we find that a transparency window appears in the probe transmission spectrum due to destructive interference. When the optical cavity is pumped on its blue sideband, the analog of electromagnetically induced absorption and parametric amplification occur due to constructive interference.
\end{abstract}

\maketitle
 \section{Introduction}
The rapidly developing research field of cavity optomechanics explores the nonlinear coupling between the electromagnetic and mechanical systems via radiation pressure, which allows for quantum control of the mechanical motion with light, and at the same time, controlling the optical response of optomechanical systems due to mechanical interactions \cite{Kippenberg2,Marquardt,Aspelmeyer}. Depending on the frequency of the electromagnetic field, the fields of optomechanics and electromechanics have witnessed remarkable progress both in optical and microwave domains, including quantum ground state cooling of the nanomechanical resonators \cite{Teufel1,Chan}, quantum coherent coupling between light and mechanical degrees of freedom \cite{Verhagen,Palomaki}, normal-mode splitting \cite{Dobrindt1,Groblacher}, and opto-/electromechanically induced transparency (OMIT/EMIT) \cite{Agarwal,Weis,Naeini,Teufel2}. OMIT/EMIT is the analog of electromagnetically induced transparency (EIT) \cite{Fleischhauer,Wu}, which has been first observed in atomic vapors \cite{Boller} and recently in various solid state systems such as quantum wells \cite{Phillips} and nitrogen-vacancy centers \cite{Santori}. Likewise, OMIT/EMIT can be utilized for slowing or advancing electromagnetic signals \cite{Naeini,Zhou} and for information storage and retrieval in the long-lived oscillations \cite{Fiore}. In addition, the phenomena of electromechanically induced absorption (EMIA) \cite{Hocke} and amplification \cite{Massel} with a blue-detuned pump field have also been demonstrated in the circuit nano-electromechanical system consisted of a superconducting microwave resonator and a nanomechanical beam. Singh \emph{et al.} have also demonstrated similar phenomena based on the optomechanical coupling between a graphene mechanical resonator and a superconducting microwave cavity \cite{Singh}.

More recently, the ability of a mechanical resonator to couple to electromagnetic cavities of vastly different wavelengths leads to intensive investigation on the double-cavity optomechanical systems. Such systems can serve as an interface to realize coherent transfer between two optical wavelengths \cite{Tian1,Tian2,Hill} and between the microwave and optical domains \cite{Barzanjeh,McGee}. In particular, by combing the technologies of electromechanics and optomechanics, Addrews \emph{et al.} have successfully coupled a nanomechanical resonator to both a superconducting microwave cavity and an optical cavity \cite{Andrews}, which consists of a hybrid opto-electromechanical system \cite{Regal,Lv}. They experimentally demonstrated reversible and coherent conversion between microwave and optical light, which can find vital applications in modern communications networks. Furthermore, Fong \emph{et al.} experimentally demonstrated coherent absorption and amplification, and the more general asymmetric Fano resonance in a cavity piezo-optomechanical system, where the optical and microwave modes are coupled to the mechanical mode through radiation pressure and piezoelectric force \cite{Fong}. In the present paper, we investigate the optical switching from the analog of electromagnetically induced transparency to absorption and amplification in the hybrid opto-electromechanical system, where both the optical and microwave modes are coupled to the mechanical mode via radiation pressure force. It should be noted that Qu and Agarwal have theoretically shown the existence of an absorption peak within the transparency window in the same double-cavity configuration when both cavities are driven at their respective red sidebands \cite{Qu}. However, the regime, where the decay rate of the optical cavity $\kappa_o$ is larger than the damping rate the mechanical resonator $\gamma_m$ that is then larger than the decay rate of the microwave cavity $\kappa_e$ (i. e., $\kappa_o>\gamma_m>\kappa_e$), should be reached to enable this phenomenon. This regime has seldom be realized in current experiment. Recently, Nunnenkamp \emph{et al.} have explored for the first time the $\emph{reversed dissipation regime}$ where the mechanical energy relaxation rate exceeds the energy decay rate of the electromagnetic cavity, which allows for mechanically-induced amplification of the electromagnetic mode \cite{Nunnenkamp}. Here, using the experimentally realizable parameters \cite{Teufel1,Teufel2,Andrews}, we show that when both cavities are pumped on their respective red sidebands, the analog of EIT appears, while the analog of electromagnetically induced absorption (EIA) \cite{Lezama} and amplification exist when one cavity is driven by a blue-detuned pump field. Therefore, we can switch conveniently from EIT to EIA and parametric amplification or vice versa by changing the frequency and power of the pump fields. This hybrid system could be utilized to realize quantum-limited amplification with unlimited gain-bandwidth product \cite{Metelmann}.

\section{Model and Theory}
The schematic of the hybrid opto-electromechanical system is shown in Fig. 1, where a microwave cavity and an optical cavity are coupled to a common nanomechanical resonator. The optical cavity with resonance frequency $\omega_o$ is driven by a strong pump beam $E_o$ with frequency $\Omega_o$ and a weak probe beam $E_p$ with frequency $\Omega_p$ simultaneously. The microwave cavity with resonance frequency $\omega_e$, denoted by equivalent inductance $L$ and equivalent capacitance $C$, is only driven by a strong pump beam $E_e$ with frequency $\Omega_e$. In a rotating frame at the pump frequency $\Omega_o$ and $\Omega_e$, the Hamiltonian of the hybrid system reads as $H=H_0+H_{\mathrm{int}}+H_{\mathrm{drive}}$, where
\begin{eqnarray}
&&H_0=\hbar\Delta_o a^\dagger a+\hbar\Delta_e b^\dagger b+\hbar\omega_m c^\dagger c,\nonumber\\
&&H_{\mathrm{int}}=-\hbar g_o a^\dagger a(c^\dagger+c)-\hbar g_e b^\dagger b(c^\dagger+c),\\
&&H_{\mathrm{drive}}=i\hbar\sqrt{\kappa_{o,ext}}E_{o}(a^\dagger -a)+i\hbar\sqrt{\kappa_{e,ext}}E_{e}(b^\dagger -b)
+i\hbar\sqrt{\kappa_{o,ext}}E_{p}(a^\dagger
e^{-i\delta t}-a e^{i\delta t}).\nonumber
\end{eqnarray}
Here, operators $a$, $b$, and $c$ are the annihilation operators of optical cavity, microwave cavity, and mechanical resonator,
respectively. $\Delta_o=\omega_o-\Omega_o$ and $\Delta_e=\omega_e-\Omega_e$ are the corresponding cavity-pump field detunings.
$\omega_m$ is the resonance frequency of the mechanical resonator with damping rate $\gamma_m$. $g_o$ $(g_e)$ is the
single-photon coupling rate between the mechanical mode and the optical (microwave) cavity mode. $H_{\mathrm{drive}}$ describes
the interaction between the input fields and the cavity fields, where $E_o$, $E_e$, and $E_p$ are related to the power of the
applied fields by $\left\vert
E_o\right\vert=\sqrt{2P_o\kappa_o/\hbar\Omega_o}$, $\left\vert
E_e\right\vert=\sqrt{2P_e\kappa_e/\hbar\Omega_e}$, and $\left\vert
E_p\right\vert=\sqrt{2P_p\kappa_o/\hbar\Omega_p}$ [$\kappa_o$ $(\kappa_e)$ is the linewidth of the optical (microwave) cavity mode], respectively. $\kappa_{o,ext}$ $(\kappa_{e,ext})$ represents the rate at which energy leaves the optical (microwave) cavity into propagating fields \cite{Andrews}. $\delta=\Omega_p-\Omega_o$ is the detuning between the probe laser beam and the pump laser beam.

According to the Heisenberg equations of motion and the commutation relation $[a,a^\dagger]=1$, $[b,b^\dagger]=1$, and $[c,c^\dagger]=1$, the temporal evolutions of operators $a$, $b$, and $Q$ [which is defined as $Q=c^\dagger+c$] can be obtained. In addition, introducing the corresponding noise and damping terms for the cavity and mechanical modes, we derive the quantum Langevin equations as follows:
\begin{eqnarray}
&\dot{a}=-i(\Delta_o-g_o Q)a-\kappa_o a+\sqrt{\kappa_{o,ext}}(E_o+E_pe^{-i\delta t})+\sqrt{2\kappa_o}a_{in},\\
&\dot{b}=-i(\Delta_e-g_e Q)b-\kappa_e b+\sqrt{\kappa_{e,ext}}E_e+\sqrt{2\kappa_e}b_{in},\\
&\ddot{Q}+\gamma_m\dot{Q}+\omega_m^2 Q=2g_o\omega_ma^\dagger a+2g_e\omega_m b^\dagger b+\xi,
\end{eqnarray}
where $a_{in}$ and $b_{in}$ are the input vacuum noise with zero mean value and $\xi$ is the Brownian stochastic force with zero mean value \cite{Genes}.
We derive the steady-state solution to Eqs. (2)-(4) by setting all the time derivatives to zero, which are given by
\begin{eqnarray}
a_{s}=\frac{\sqrt{\kappa_{o,ext}}E_o}{\kappa_o+i\Delta_o'}, b_{s}=\frac{\sqrt{\kappa_{e,ext}}E_e}{\kappa_e+i\Delta_e'},
Q_s=\frac{2}{\omega_m}(g_o\left\vert a_{s}\right\vert^2+g_e\left\vert b_{s}\right\vert^2),
\end{eqnarray}
where $\Delta_o'=\Delta_o-g_o Q_s$ and $\Delta_e'=\Delta_e-g_e Q_s$ are the effective cavity detunings including radiation pressure effects.
Subsequently, we rewrite each Heisenberg operator as the sum of its steady-state mean value and a small fluctuation with zero mean value (i.e., $a=a_{s}+\delta a, b=b_{s}+\delta b, Q=Q_s+\delta Q$) and obtain the following linearized Heisenberg-Langevin equations:
\begin{eqnarray}
&\left\langle\delta\dot{a}\right\rangle=-(\kappa_o+i\Delta_o)\left\langle\delta a\right\rangle+ig_o Q_s\left\langle\delta a\right\rangle+ig_o a_{s}\left\langle\delta Q\right\rangle+\sqrt{\kappa_{o,ext}}E_p e^{-i\delta t},\\
&\langle\delta\dot{b}\rangle=-(\kappa_e+i\Delta_e)\left\langle\delta b\right\rangle+ig_e Q_s\left\langle\delta b\right\rangle+ig_e b_{s}\left\langle\delta Q\right\rangle,\\
&\langle\delta\ddot{Q}\rangle+\gamma_m\langle\delta\dot{Q}\rangle+\omega_m^2\langle\delta Q\rangle=2\omega_m g_o a_{s}(\langle\delta a\rangle+\langle\delta a^\dagger\rangle)+2\omega_m g_e b_{s}(\langle\delta b\rangle+\langle\delta b^\dagger\rangle).
\end{eqnarray}
where we have identified all operators with their expectation values and dropped the quantum and thermal noise terms\cite{Weis}. The nonlinear terms $\delta a^\dagger\delta a$, $\delta b^\dagger\delta b$, $\delta a\delta Q$, and $\delta b\delta Q$ can result in some interesting phenomena of optomechanical systems, such as second and higher-order sideband \cite{Xiong}. When the pump power is strong enough, $\left\vert a_{s}\right\vert\gg1$ and $\left\vert b_{s}\right\vert\gg1$, therefore, one can safely neglect these nonlinear terms in the linearized equations (6)-(8) \cite{Genes}.
In order to solve equations (6)-(8), we
make the ansatz \cite{Boyd} $\langle\delta a\rangle=a_{+}e^{-i\delta
t}+a_{-}e^{i\delta t}$, $\langle\delta b\rangle=b_{+}e^{-i\delta
t}+b_{-}e^{i\delta t}$, and $\langle\delta Q\rangle=Q_+e^{-i\delta
t}+Q_-e^{i\delta t}$. Upon substituting the above ansatz into Eqs. (6)-(8), we can obtain the following solution
\begin{eqnarray}
a_{+}=\frac{\sqrt{\kappa_{o,ext}}E_p}{\kappa_o+i\Delta_o'-i\delta}-\frac{ig_o^2 n_{o}}{f(\delta)}\frac{\sqrt{\kappa_{o,ext}}E_p}{(\kappa_o+i\Delta_o'-i\delta)^2},
\end{eqnarray}
where
\begin{eqnarray}f(\delta)=\frac{2\Delta_o'g_o^2n_{o}}{(\kappa_o-i\delta)^2+\Delta_o'^2}+\frac{2\Delta_e'g_e^2n_{e}}{(\kappa_e-i\delta)^2+\Delta_e'^2}
-\frac{\omega_m^2-\delta^2-i\delta\gamma_m}{\omega_m}.
\end{eqnarray}
Here, $n_{o}=\left\vert a_{s}\right\vert^2$ and $n_{e}=\left\vert b_{s}\right\vert^2$, approximately equal to the number of
pump photons in each cavity, are determined by the following coupled equations
\begin{eqnarray}
n_{o}=\frac{\kappa_{o,ext}E_o^2}{\kappa_o^2+\left[\Delta_o-2g_o/\omega_m(g_o n_{o}+g_e n_{e})\right]^2},\\
n_{e}=\frac{\kappa_{e,ext}E_e^2}{\kappa_e^2+\left[\Delta_e-2g_e/\omega_m(g_o n_{o}+g_e n_{e})\right]^2}.
\end{eqnarray}
The output field from the cavity can be obtained by using the standard
input-output theory \cite{Gardiner}
$a_{out}(t)=a_{in}(t)-\sqrt{\kappa_{ext}}a(t)$, where $a_{out}(t)$ is the
output field operator. Considering the output field of the optical cavity, we have
\begin{eqnarray}
\left\langle a_{out}(t)\right\rangle&=&(E_o-\sqrt{\kappa_{o,ext}}a_{s})e^{-i\Omega_o t}+(E_p-\sqrt{\kappa_{o,ext}}a_{+})
e^{-i(\delta+\Omega_o)t}-\sqrt{\kappa_{o,ext}}a_{-}e^{i(\delta-\Omega_o)t}\nonumber\\
&=&(E_o-\sqrt{\kappa_{o,ext}}a_{s})e^{-i\Omega_o t}+(E_p-\sqrt{\kappa_{o,ext}}a_{+})
e^{-i\Omega_p t}-\sqrt{\kappa_{o,ext}}a_{-}e^{-i(2\Omega_o-\Omega_p)t}.\nonumber\\
\end{eqnarray}
Defining the transmission of the probe field as the ratio of the
output and input field amplitudes at the probe frequency \cite{Weis}, we have
\begin{eqnarray}
t(\Omega_p)&=&\frac{E_p-\sqrt{\kappa_{o,ext}}a_{+}}{E_p}\nonumber\\&=&1-\left[\frac{\kappa_{o,ext}}
{\kappa_o+i\Delta_o'-i\delta}-\frac{1}{f(\delta)}\frac{ig_o^2 n_{o}\kappa_{o,ext}}{(\kappa_o+i\Delta_o'-i\delta)^2}\right].
\end{eqnarray}
The tunable probe transmission window will modify the propagation dynamics of a probe field sent to this hybrid system due to the
variation of the complex phase picked by its different frequency components. The probe laser beam will experience a group delay
$\tau_{g}$ expressed as
\begin{eqnarray}
\tau_{g}=\left.\frac{d\phi}{d\Omega_{p}}\right|_{\Omega_p=\omega_{o}},
\end{eqnarray}
where $\phi=\arg[t(\Omega_p)]$ is the rapid phase dispersion.

\section{Numerical results and discussion}
In what follows, we consider for illustration an experimentally realizable hybrid opto-electromechanical system. The parameters used are \cite{Teufel1,Teufel2,Andrews}: $\omega_o=2\pi\times282$ THz, $\omega_e=2\pi\times7.1$ GHz, $\kappa_o=2\pi\times1.65$ MHz, $\kappa_e=2\pi\times1.6$ MHz, $\kappa_{o,ext}=0.76\kappa_o$, $\kappa_{e,ext}=0.11\kappa_e$, $g_o=2\pi\times27$ Hz, $g_e=2\pi\times2.7$ Hz, $\omega_m=2\pi\times5.6$ MHz, and $\gamma_m=2\pi\times4$ Hz.

Optical response of this hybrid system is characterized by the transmission of the probe laser beam under the influence of two pump beams. Firstly, we consider the situation where both optical and microwave cavities are pumped on their respective red sidebands, i.e., $\Delta_o=\Delta_e=\omega_m$. In our previous work, we have investigated EIT and slow light in a two-mode optomechanical system when two optical cavities are driven by red detuned pump beams \cite{Jiang}. Figure 2 plots the magnitude of transmission and phase dispersion of the transmitted probe beam as a function of the probe-cavity detuning $\Delta_p$ for $P_o=0$, 2 and 3 mW, respectively. From Fig. 2(a), we can see that there is a transmission dip in the center of the probe transmission spectrum when the optical pump power $P_o=0$. However, this broad cavity resonance can be split into two dips in the presence of the pump laser beam and a significant transparency window appears at the resonant region, i.e., $\Delta_p=0$. Moreover, the transparency window can be effectively modulated by the pump laser, which has been demonstrated by Weis \cite{Weis}, Safavi-Naeini \cite{Naeini}, and Teufel \cite{Teufel2} in the typical single-mode optomechanical systems in both optical and microwave domains. The underlying physical mechanism for this phenomenon can be explained as follows. The simultaneous presence of pump and probe fields induces a modulation at the beat frequency $\delta=\Omega_p-\Omega_o$ of the radiation pressure force acting on the common mechanical resonator. When this modulation is close to the mechanical resonance frequency $\omega_m$, the mechanical mode starts to oscillate coherently, giving rise to Stokes ($\Omega_s=\Omega_o-\omega_m$) and anti-Stokes ($\Omega_{as}=\Omega_o+\omega_m$) scattering of light from the strong pump field. When the optical cavity is driven on its red sideband, the highly off-resonant Stokes scattering is suppressed and only the anti-Stokes scattering builds up within the cavity. However, the near-resonant probe beam ($\Delta_p\approx0$) can interfere with the anti-Stokes field destructively, and as a result, the probe transmission spectrum can be modified due to mechanical vibration. The analog of EIT effect in this hybrid system can also cause an extremely steep and positive phase shift for the transmitted probe beam, as shown in Fig. 2(b). The derivative of such a phase shift leads to the group delay due to the causality-preserving subluminal effects where a probe pulse contained within the transparency window would accumulate in its transmission through the cavity. The maximum signal time delay can be calculated according to Eq. (15) when pump-probe detuning $\delta$ is equal to $\omega_m$. Similar to Ref. \cite{Naeini}, the reflected probe beam would experience tunable group advance. Such effects have been investigated in detail by us in two-mode optomechanical systems \cite{Jiang}. However, the hybrid system we study here can also be used to slow and advance the microwave signals \cite{Zhou}.

We have then considered the situation where the optical cavity mode is pumped on its blue sideband while the microwave cavity is pumped on its redsideband, i.e., $\Delta_o=-\omega_m$ and $\Delta_e=\omega_m$. In Fig. 3, the probe transmission $|t|^2$ is plotted as a function of the probe-cavity detuning $\Delta_p$ for $P_o=$ 0, 10, and 40 $\mu$W, respectively. When the pump field is off, the probe transmission spectrum shows the usual Lorentzian line shape of the bare cavity, which can be seen from Fig. 3(a). However, when $P_o=$10 $\mu$W, we can see that the cavity transmission is reduced in the resonant region compared with bare cavity minimum ($P_o=0$), which is referred to as the analog of EIA. At even higher pump power, $P_o=40$ $\mu$W for example, the system switches from EIA \cite{Naeini,Hocke} to parametric amplification \cite{Massel}, where the probe transmission can be amplified significantly. In this case, the hybrid system can be used as a transistor \cite{Chen} to amplify the weak optical or microwave signal. These phenomena can be understood in terms of constructive interference. When the optical cavity is driven by a blue-detuned pump beam, the anti-Stokes scattering is off-resonant with the cavity and therefore is strongly suppressed, whereas the Stokes scattering builds up within the cavity. Constructive interference between the Stokes field and the near-resonant probe field sent to the cavity enhances the build-up of the intra-cavity probe field. The resulting increased feeding of probe photons into the cavity manifests itself as a reduced cavity transmission. When the pump power is strong enough, the number of down-converted pump photons is much larger than the probe photons sent to the cavity, and the probe transmission can exceed unity because of the parametric amplification of the probe beam by the hybrid opto-electromechanical system. The peak probe transmission $|t|^2$ versus the pump power is plotted in Fig. 4. It can be seen clearly that the peak probe transmission decreases from a value for the bare cavity minimum ($P_o=0$) to a minimum value with the increasing pump power, representing the analog of EIA. Increasing the pump power further, the probe transmission begins to increase gradually and exceeds the initial value for the bare cavity. When the pump power is above about 37 $\mu$W, $|t|^2>1$, the hybrid system enters the regime of parametric amplification, where the probe beam can be amplified significantly. Based on the above discussions, we can switch conveniently from the analog of EIT to EIA and parametric amplification by adjusting the frequency and power of the pump field. Such a phenomenon should also exist when the weak probe beam is applied to the microwave cavity.

\section{Conclusion}
In conclusion, we have studied the mechanically-mediated optical response of a hybrid opto-electromechanical system consisted of an optical cavity and a microwave cavity coupled to a common mechanical resonator. When both cavities are pumped on their respective red sidebands, the analog of electromagnetically induced transparency occurs due to destructive interference between the generated anti-Stokes field and the probe field. However, when the optical cavity is driven by a blue-detuned pump field, constructive interference between the Stokes filed and the probe field leads to the analog of electromagnetically induced absorption and parametric amplification. These phenomena can be switched from one to another easily by controlling the frequency and power of the pump fields.

\section{acknowledgments} The authors gratefully acknowledge support
from National Natural Science Foundation of China (Grant Nos. 11304110 and 11174101), Jiangsu Natural Science Foundation (Grant Nos. BK20130413 and BK2011411), Natural Science Foundation of the Jiangsu Higher Education Institutions of China (Grant No. 13KJB140002), and Science and Technology support program of Huaian (HAG2011006).

\clearpage
\begin{figure}
\includegraphics[width=12cm]{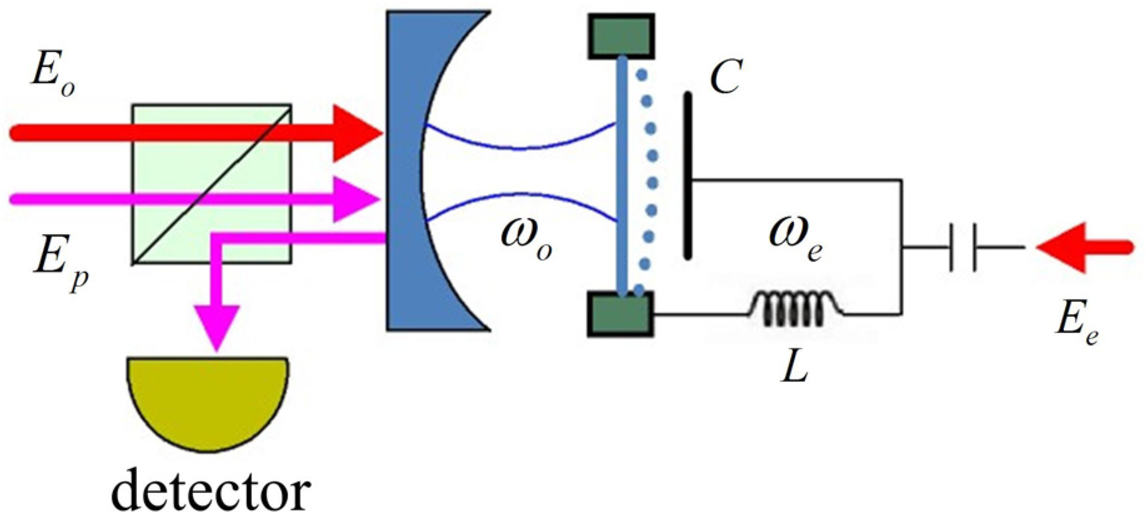}
\caption{Schematic of a two-mode optomechanical system where two optical cavity modes, $a_1$ and $a_2$, are coupled to the same mechanical mode $b$. The left cavity is driven by a strong pump beam $E_L$ in the simultaneous presence of a weak probe beam $E_p$ while the right cavity is only driven by a pump beam $E_R$.}
\end{figure}

\clearpage
\begin{figure}
\includegraphics[width=12cm]{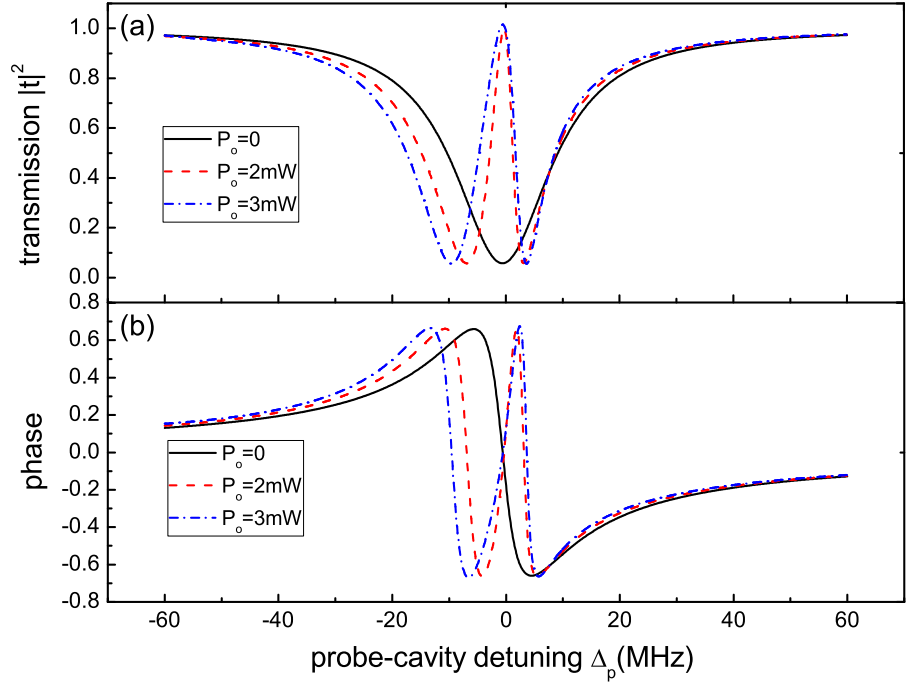}
\caption{(a) Magnitude and (b) phase of the transmitted probe beam as a function of the probe-cavity detuning $\Delta_p=\Omega_p-\omega_o$ for optical pump power $P_o$ equal to 0, 2, and 3 mW, respectively. The right pump power is kept equal to 1 $\mu$W. Both cavities are pumped on their respective red sidebands, i.e., $\Delta_o=\omega_m$ and $\Delta_e=\omega_m$.
The other parameters used are $\omega_o=2\pi\times282$ THz, $\omega_e=2\pi\times7.1$ GHz, $\kappa_o=2\pi\times1.65$ MHz, $\kappa_e=2\pi\times1.6$ MHz, $\kappa_{o,ext}=0.76\kappa_o$, $\kappa_{e,ext}=0.11\kappa_e$, $g_o=2\pi\times27$ Hz, $g_e=2\pi\times2.7$ Hz, $\omega_m=2\pi\times5.6$ MHz, $\gamma_m=2\pi\times4$ Hz.}
\end{figure}

\clearpage
\begin{figure}
\centering
\includegraphics[width=12cm]{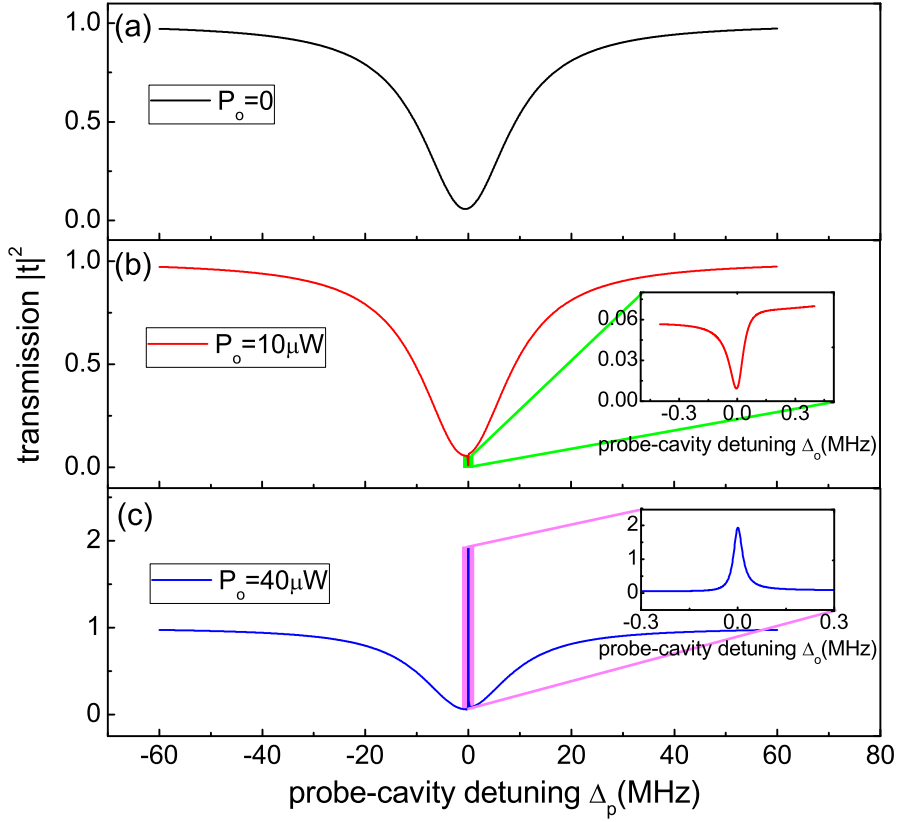}
\caption{Probe transmission $|t|^2$ versus probe-cavity detuning $\Delta_p$ for $P_o=0$, 10, and 40 $\mu$W, respectively. Here, the optical cavity is pumped on its blue sideband and the microwave cavity is pumped on its red sideband, i.e., $\Delta_o=-\omega_m$ and $\Delta_e=\omega_m$. The inset of Fig. 3(b) and Fig. 3(c) show the probe transmission near the resonant region on an enlarged scale. Other parameters are $P_e=1 $ $\mu$W, $\omega_o=2\pi\times282$ THz, $\omega_e=2\pi\times7.1$ GHz, $\kappa_o=2\pi\times1.65$ MHz, $\kappa_e=2\pi\times1.6$ MHz, $\kappa_{o,ext}=0.76\kappa_o$, $\kappa_{e,ext}=0.11\kappa_e$, $g_o=2\pi\times27$ Hz, $g_e=2\pi\times2.7$ Hz, $\omega_m=2\pi\times5.6$ MHz, $\gamma_m=2\pi\times4$ Hz.}
\end{figure}

\clearpage
\begin{figure}
\centering
\includegraphics[width=12cm]{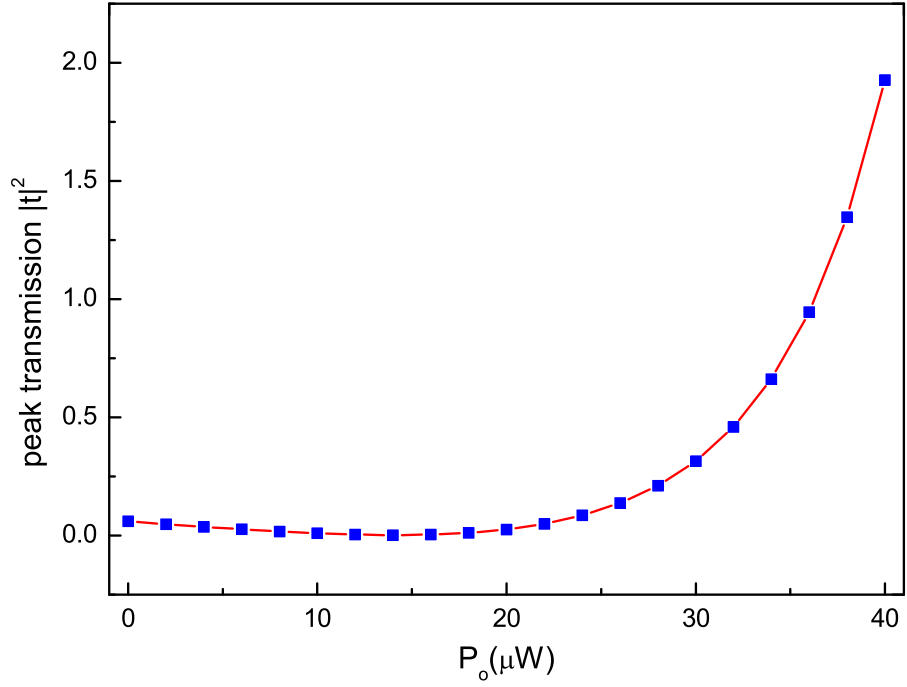}
\caption{Peak probe transmission $|t|^2$ at the cavity resonance as a function of the optical pump power $P_o$ when $\Delta_o=-\omega_m$ and $\Delta_e=\omega_m$. Other parameters are the same with those of figure 3.}
\end{figure}

\end{document}